\documentclass[aps,twocolumn,pra,reprint,amsmath,amssymb,floatfix,footinbib,superscriptaddress]{revtex4}

\usepackage{amsmath}	
\usepackage{gensymb}
\usepackage{hyperref}
\hypersetup{colorlinks=true}
\usepackage{upgreek}
\usepackage{graphics}
\usepackage{hyperref}
\usepackage{epsfig}
\usepackage{color}
\usepackage{bm}
\usepackage{float}
\usepackage{ulem} 

\usepackage{epstopdf}

\usepackage{CJK}        
\usepackage{url}
\usepackage{multirow}

\usepackage{xcolor}

\newcommand{\upperRomannumeral}[1]{\uppercase\expandafter{\romannumeral#1}}

\begin{document}
\begin{CJK*}{GBK}{kai}

\title{Nontrivial criticality and weakly first-order transitions in the frustrated spin-1 chain with single-ion anisotropy}

\author{Qiang Luo}
\affiliation{College of Physics, Nanjing University of Aeronautics and Astronautics, Nanjing, 211106, China}
\affiliation{School of Physics, Renmin University of China, Beijing 100872, China}

\author{Shijie Hu}
\email[]{shijiehu@csrc.ac.cn}
\affiliation{Beijing Computational Science Research Center, Beijing 100084, China}
\affiliation{Physics Department and Research Center OPTIMAS, RPTU University Kaiserslautern-Landau, 67663 Kaiserslautern, Germany}

\author{Jize Zhao}
\affiliation{School of Physical Science and Technology $\&$ Key Laboratory of Quantum Theory and Applications of MoE, Lanzhou University, Lanzhou 730000, China}
\affiliation{Lanzhou Center for Theoretical Physics, Key Laboratory of Theoretical Physics of Gansu Province, Lanzhou University, Lanzhou 730000, China}

\author{Sebastian Eggert}
\email[]{seggert@rptu.de}
\affiliation{Physics Department and Research Center OPTIMAS, RPTU University Kaiserslautern-Landau, 67663 Kaiserslautern, Germany}

\author{Xiaoqun Wang}
\email[]{xiaoqunwang@zju.edu.cn}
\affiliation{School of Physics and Institute for Advanced Study in Physics, Zhejiang University, Hangzhou 310058, China}
\affiliation{School of Physics and Astronomy, Tsung-Dao Lee Institute, Shanghai Jiao Tong University, Shanghai 200240, China}

\date{\today}

\begin{abstract}
 Frustrated spin chains host exotic phases beyond the Landau paradigm, yet weak first-order transitions and unconventional criticalities remain elusive. Combining large-scale density-matrix renormalization group with bosonization methods, we map the phase diagram of a spin-1 chain with next-nearest-neighbor (NNN) coupling and single-ion anisotropy (SIA). We pinpoint two topological phase transitions: a weakly first-order transition between the Haldane and NNN-Haldane phases with a tiny gap and a continuous transition from the Haldane phase to the large-$D$ phase located via level spectroscopy with Richardson extrapolation. For negative SIA, a spontaneous dimerized phase emerges between the N\'eel and NNN-N\'eel phases. Its transition into the N\'eel phase is a deconfined quantum critical point falling into the eight-vertex universality class, with central charge $c = 1$ and critical exponents varying continuously along the phase boundary. For positive SIA, we identify two vector-spin-chirality orders---an ordinary chiral phase and a narrow Haldane-chiral phase---with a weakly first-order transition between them, evidenced by a soft jump in the string order parameter. Our results reveal that frustration and anisotropy generate emergent orders and non-Landau criticalities, providing a platform for deconfined and topological phase transitions.
\end{abstract}


\maketitle

\section{Introduction}
In one-dimensional quantum spin chains \cite{Giamarchi2003,Gao2026SCPMA}, the intricate interplay between competing interactions and magnetic anisotropies, notably the Dzyaloshinskii-Moriya interaction \cite{Zhao2005PRL}, bond-directional exchanges \cite{Yang2020PRL,Yang2021PRB,Luo2021PRB,Luo2021PRR}, and single-ion anisotropy (SIA) \cite{Chen2003PRB,Hu2011PRB,Tzeng2017PRB}, stabilizes a variety of exotic quantum phenomena. These include the rank-2 vector chiral and spin-nematic phases \cite{Kolezhuk2000PRB,McCulloch2008PRB}, the so-called ``half fire, half ice" phases emerging in ferrimagnets \cite{Yin2024PRL}, and the field-induced magnetization plateaus \cite{Guo2024PRB}.
Substantial advances in understanding such unusual phases are driven by a synergistic combination of theoretical, computational, and experimental progress. On the one hand, the refinement of sophisticated analytical and numerical techniques has enabled the accurate detection and characterization of subtle quantum orders. On the other hand, the continuous discovery of pertinent material realizations provides essential experimental benchmarks and poses new challenges to theory \cite{Orlova2017PRL,Li2025PRB,Vaidya2026PRB}. One prominent example is the candidate spin-nematic phase in LiCuVO$_4$, as evidenced by high-field magnetization and NMR experiments \cite{Orlova2017PRL}; another is the chiral antiferromagnetic order recently uncovered in the $S=1$ chain [Ni(pym)(H$_2$O)$_4$]SO$_4\cdot$H$_2$O, where $\mu$SR and neutron diffraction reveal a chiral modulation of the easy-axis anisotropy as the origin of chirality \cite{Vaidya2026PRB}.

\begin{figure}[!t]
  \centering
  \includegraphics[width=0.90\columnwidth, clip]{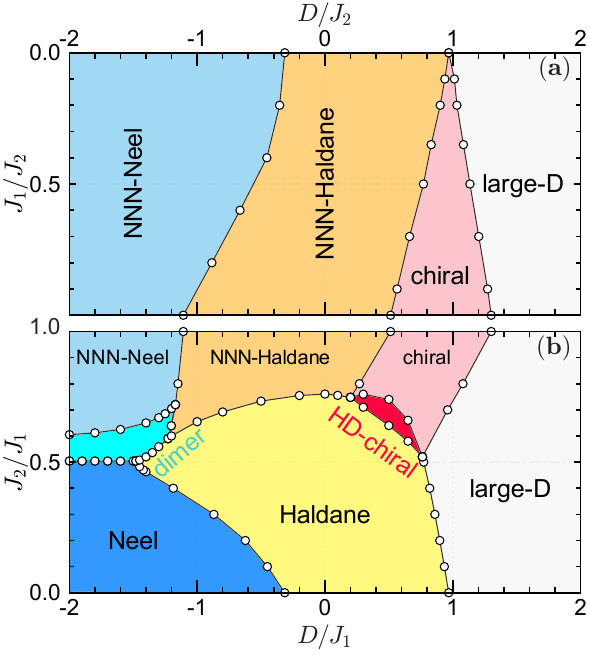}\\
  \caption{Ground-state phase diagram of the spin-1 $J_1$-$J_2$-$D$ chain in the (a) ($D/J_2$, $J_1/J_2$) plane and (b) ($D/J_1$, $J_2/J_1$) plane, respectively. The open circles represent the transition points determined by the DMRG or iDMRG calculation.}
  \label{FIG:GSPD}
\end{figure}

Spin-chain models serve as a fertile theoretical playground for exploring unconventional quantum phase transitions (QPTs) that lie beyond the conventional Landau-Ginzburg-Wilson (LGW) paradigm. Among these, two particularly prominent types have attracted sustained theoretical attention: topological QPTs involving non-Landau ordered phases, and
Landau-forbidden continuous QPTs that separate two distinct symmetry-breaking orders.
A paradigmatic example of the former is the transition into the Haldane phase \cite{Haldane1983PRL} --- a prototypical symmetry-protected topological (SPT) phase \cite{Pollmann2012PRB} characterized by a bulk gap, spin-$1/2$ edge modes, and a nonlocal string order parameter (SOP) originating from hidden $\mathbb{Z}_2\times\mathbb{Z}_2$ symmetry breaking \cite{denNijs1989PRB,Kennedy1992PRB}. The exactly solvable Affleck-Kennedy-Lieb-Tasaki (AKLT) model provides a crucial reference point for understanding its essential properties \cite{AKLT1987PRL}. Beyond this well-known SPT example, topological QPTs have been identified in various other settings where conventional order parameters are absent, including spin-$1/2$ dimerized chains, spin-1 chains with SIA \cite{Chen2003PRB,Hu2011PRB,Tzeng2017PRB,Ren2018}, frustrated spin-1 $J_1$-$J_2$ Heisenberg chains \cite{Pati1996EPL,Kolezhuk1996PRL,Kolezhuk1997PRB,Kolezhuk2002PRB}, and spin-2 chains constructed from projection operators \cite{Zang2010PRB}.
The second type---deconfined quantum criticality---is exemplified by continuous transitions between two conventionally ordered phases that would typically be separated by a first-order transition within the LGW framework \cite{Senthil2004Science,Chen2026SCPMA,Jiang2019PRB,Roberts2019PRB,Mudry2019PRB}. Such deconfined quantum critical points (DQCPs) have been proposed, for instance, in certain frustrated ferromagnetic spin chains \cite{Huang2019PRB,Sun2019PRB,Luo2019PRB} and in Kitaev-type chains supplemented with multiple-spin interactions or SIA \cite{Macedo2022PRB,Luo2023PRB}, where the transition is driven by the condensation of fractionalized excitations.

Over the years, extensive research has been devoted to spin-1 chains with competing interactions and magnetic anisotropies in order to elucidate the nature of emergent phases and characterize the corresponding QPTs.
These models include the $J_1$-$J_2$ chain with an XXZ-type exchange anisotropy \cite{Hikihara2000JPSJ} or multiple-spin interaction \cite{Pixley2014PRB,Chepiga_2016_1,Chepiga_2016_2}, 
the bilinear-biquadratic chain with SIA \cite{Rodriguez2011PRL,Chiara2011PRB} or Kitaev-type exchange anisotropy \cite{Wei2025PRB}, the Heisenberg-Kitaev chain with SIA \cite{Reja2026PRB}, and the inherently anisotropic Kitaev-$\Gamma$ chain \cite{Luo2021PRR,Luo2026arXiv}, where Kitaev and symmetric off-diagonal $\Gamma$ terms compete directly.
While the frustrated $J_1$-$J_2$ chain and the Heisenberg chain with SIA serve as two paradigmatic models for investigating competing exchange and local anisotropy separately, their synthesis---the $J_1$-$J_2$-$D$ chain---has only been explored in limited parameter regimes \cite{Hikihara2002JPSJ,Riviere2026arXiv}. 
As we shall demonstrate, the phase diagram harbors a far richer landscape than previously anticipated, including a weak first-order transition separating two distinct chiral phases, a DQCP of the eight-vertex universality class, and a spontaneous dimerized phase. It is therefore the objective of this work to present a comprehensive and high-precision phase diagram, characterizing every transition by its universality class and critical exponents, thereby providing a definitive reference for future studies on frustrated spin chains.

The remainder of this paper is organized as follows. In Sec.~\ref{SEC:Model}, we introduce the Hamiltonian of the $J_1$-$J_2$-$D$ chain, outline the numerical methods employed, and present the resulting quantum phase diagram. Section~\ref{SEC:QPT} describes our protocol for pinpointing quantum critical points with high accuracy. In Sec.~\ref{SEC:Dimer}, we focus on the negative-SIA regime, where a spontaneous dimerized phase emerges; we demonstrate that its transition to the N\'eel phase constitutes a DQCP of the eight-vertex type. Section~\ref{SEC:Chiral} analyzes the two chiral phases for positive SIA: an ordinary chiral phase and a narrow Haldane-chiral phase.
For small $J_1$ we invoke bosonization to understand the ordinary chiral phase, which is numerically shown  to undergo a  weakly first-order transition to the Haldane chiral phase.
Finally, Sec.~\ref{SEC:CONC} summarizes our findings and discusses open questions for future study.

\section{Model and Phase Diagram}\label{SEC:Model}

The Hamiltonian of the spin-1 $J_1$-$J_2$-$D$ chain is
\begin{align}\label{EQ:J1J2Dz-Ham}
\mathcal{H}
= \sum_{i} (J_1 \mathbf{S}_i\cdot\mathbf{S}_{i+1} + J_2 \mathbf{S}_i\cdot\mathbf{S}_{i+2})
+ D\sum_{i}(S_i^z)^2
\end{align}
in which $\mathbf{S}_i = (S_i^x, S_i^y, S_i^z)$ denotes the spin operator at site $i$ of a chain of total length $L$.
The parameters $J_1$ and $J_2$ are the nearest-neighbor~(NN) and next-nearest-neighbor~(NNN) Heisenberg couplings, respectively, while $D$ stands for the SIA,  which may take either positive or negative values.
Since the primary objective of this study is to discover emergent phases and identify unconventional QPTs, we systematically employ both the density-matrix renormalization group (DMRG) \cite{White1992PRL,Peschel_1999,Schollwoeck_2005} and infinite DMRG (iDMRG) methods. In the DMRG simulations, we adopt different boundary conditions as appropriate for distinct purposes: open boundary conditions (OBC) are primarily used for finite-size scaling (FSS) analysis, with retained states kept up to 2000 and sweeps performed up to 12 times to ensure convergence; periodic (PBC) and twisted boundary conditions (TBC), on the other hand, are employed to mitigate finite-size effects and thereby accurately determine quantum critical points.

In the following sections we analyze the location and the nature of the phase transition lines in detail. The resulting ground-state phase diagram of Eq.~\eqref{EQ:J1J2Dz-Ham} is summarized in Fig.~\ref{FIG:GSPD}. Panels (a) and (b) correspond to the parameter ranges $0 \!\le\! J_1/J_2\! \le \!1$ and $0 \!\le J_2/J_1\! \le 1$, respectively. In total, eight distinct phases are identified. Without frustration $J_2\!=\!0$ there are three known phases:  an Ising-like ordered N\'eel phase for predominantly negative $D$, a Haldane phase and a large-$D$ phase above a topological transition at  $D_c \!=\! 0.96847256(2)$, which we determined as
a benchmark value for the critical point. For $J_1\!=\!0$ the system is described by two separate chains with in-chain coupling $J_2$, so the same phases appear again with the correlations shifted from nearest neighbor to effectively next-nearest neighbor order (see Fig.~\ref{FIG:GSPD}(a)).  
However, a finite $J_1$ induces an interesting chiral order \cite{Okunishi2008,He2024PRB} close to $D_c$ in this case, which will be analyzed using bosonization below.  At intermediate values of frustration the competition of $J_1$ and $J_2$ leads to the emergence of two more phases: a spontaneously dimerized phase and a Haldane-chiral phase with nonzero vector spin chirality \cite{Okunishi2008,He2024PRB}.
Notably, two intriguing aspects of the QPTs are observed. First, the transition between the N\'eel phase and the dimerized phase is continuous with central charge $c=1$ and falls into the eight-vertex universality class. This transition lies outside the LGW paradigm, as the broken symmetries---time reversal and translation---are unrelated. Second, with increasing SIA, the transition between the conventional chiral phase and the Haldane-chiral phase is identified as weakly first order, as signaled by a soft jump in the SOP, with its weak nature reflecting the near-degeneracy of competing chiral and topological orders.

\section{Trilogy of Quantum Phase Transitions}\label{SEC:QPT}

\subsection{Haldane--NNN-Haldane weakly first-order transition}

We begin by examining the QPT in the frustrated spin-$1$ $J_1$-$J_2$ chain without SIA ($D = 0$) \cite{Pati1996EPL,Kolezhuk1996PRL,Kolezhuk1997PRB,Kolezhuk2002PRB}.
In the frustration-free limit $J_2/J_1 = 0$, the ground state resides in the well-known Haldane phase, characterized by a nonlocal SOP \cite{denNijs1989PRB,Kennedy1992PRB}. This phase hosts a bulk gap $\Delta_H$ = 0.41050(2) (see Refs.~\cite{Ejima2015,Fang2022FrontPhys} and references therein) and supports two spin-$1/2$ edge states \cite{Pollmann2012PRB}, which give rise to a quasi-fourfold ground-state degeneracy in OBC. These edge states interact via an effective coupling that decays exponentially with chain length.
At weak frustration, the Haldane phase remains the dominant ground state. Notably, Chepiga and Mila demonstrated numerically that the effective coupling between the fractionalized edge states in the Haldane phase can change sign repeatedly as $J_2/J_1$ increases \cite{ChepigaMila_2017}. This sign reversal leads to a proliferation of exact zero modes, whose number scales linearly with the chain length.
Upon further increasing frustration, the system undergoes a weakly first-order QPT. However, unlike the extensively studied spin-1/2 counterpart \cite{Okamoto_1992,Castilla_1995,Eggert_1996,White1996PRB}, frustration here does not induce dimerization.
Rather, it stabilizes a so-called NNN-Haldane phase. This phase is topologically trivial and can be characterized by an alternative nonlocal SOP composed of two intertwined strings \cite{Kolezhuk1996PRL,Kolezhuk1997PRB,Kolezhuk2002PRB}.
Exploiting the exactly known AKLT state as the trial wave function \cite{AKLT1987PRL}, it is estimated that the transition point $J_{2,t}^{\textrm{var}}/J_1$ = $1/(1+e^{-1/\xi})$ = $3/4$ (see Supporting Information \cite{SuppMat}) where $\xi = 1/{\ln3}$ is the correlation length at the AKLT point \cite{AKLT1987PRL}.

\begin{figure}[!t]
  \centering
  \includegraphics[width=0.90\columnwidth, clip]{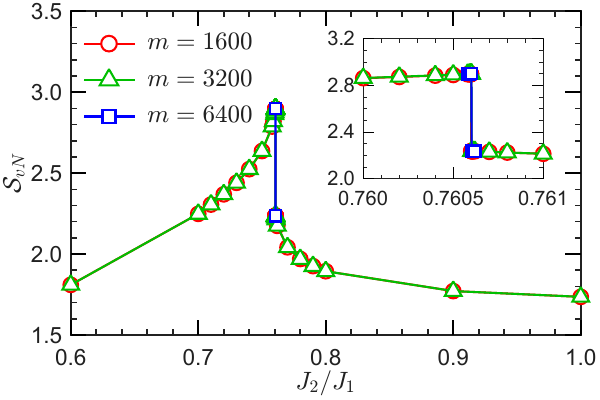}\\
  \caption{The entanglement entropy $\mathcal{S}_{vN}$ versus $J_2/J_1$ in the $J_1$-$J_2$ chain. The truncated dimension $m$ kept in the iDMRG calculation is 1600 (red circles), 3200 (green triangles), and 6400 (blue squares). The inset magnifies $\mathcal{S}_{vN}$ in the region where $0.760 \leq J_2/J_1 \leq 0.761$.}
  \label{FIG:J1J2VNE}
\end{figure}

Nevertheless, a high-precision, numerically substantiated benchmark for this critical point has remained elusive, largely due to the weakly first-order nature of the transition. Early DMRG studies by Kolezhuk \textit{et al.} placed the transition point $J_{2,t}/J_1$ between 0.7444(6) \cite{Kolezhuk1996PRL} and 0.77 \cite{Kolezhuk2002PRB}, based on discontinuities in these two nonlocal SOPs, while the precise location of the transition within this range was left unresolved. More recently, Vanderstraeten \textit{et al.} quoted an improved estimate of $J_{2,t}/J_1 \approx 0.7606$ \cite{Vanderstraeten2020}; however, the numerical details supporting this value were not elaborated, leaving its reliability and uncertainty unquantified. This persistent ambiguity highlights the need for a systematic and transparent numerical determination.

As a representative example of weakly first-order QPTs, the transition point cannot be accurately determined by the singularity in the energy curve.
Since the Haldane phase and the NNN-Haldane phase are composed of dominant NN and NNN valence bonds, respectively, they can be distinguished by their entanglement properties.
By now it is well established that entanglement provides a unique framework for capturing QPTs and critical phenomena in quantum many-body systems \cite{Vidal2003PRL,Calabrese2004JSTAT}.
Particularly, the von Neumann entanglement entropy is defined as $\mathcal{S}_{vN}(L, l) = -\mathrm{Tr}(\rho_l\ln\rho_l)$, where $\rho_l$ is the reduced density matrix of a given subsystem of length $l$ embedded in a whole system of length $L$.
In the conformal invariant critical region, the entanglement entropy for a finite system obeys the scaling behavior
\begin{equation}\label{EQ:VNE}
\mathcal{S}_{vN}(L, l) = \frac{c}{6}\ln \left(\frac{2L}{\pi}\sin \frac{\pi l}{L}\right) + \mathcal{S}' + \cdots,
\end{equation}
where $c$ is the central charge, $\mathcal{S}'$ is a model-dependent nonuniversal constant, and the ellipsis represents other correlations such as the alternating term due to boundary effect \cite{Vidal2003PRL,Calabrese2004JSTAT}.
In iDMRG simulations at criticality, the finite bond dimension $m$ introduces an effective finite correlation length $\xi(m) \sim m^{\kappa}$, where $\kappa$ is related to the central charge via $\kappa = 6/(c\sqrt{12/c+1})$. This makes the entanglement entropy scale with $m$ as $\mathcal{S}_{vN} \simeq \ln m/\sqrt{12/c+1}$ \cite{Pollmann2009PRL}.

\begin{figure}[!t]
\centering
\includegraphics[width=0.90\columnwidth, clip]{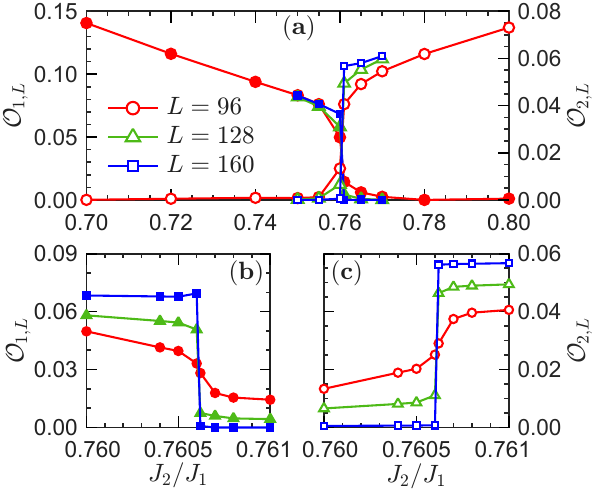}\\
  \caption{(a) The SOPs $\mathcal{O}_{1,L}$ (left axis, for the Haldane phase) and $\mathcal{O}_{2,L}$ (right axis, for the NNN-Haldane phase) versus $J_2/J_1$ for the $J_1$-$J_2$ chain. Data are shown for chain lengths $L = 96$ (red circles), $128$ (green triangles), and $160$ (blue squares). Panels (b) and (c) provide magnified views of $\mathcal{O}_{1,L}$ and $\mathcal{O}_{2,L}$, respectively, within the interval $0.760 \leq J_2/J_1 \leq 0.761$.}
\label{FIG:J1J2SOP}
\end{figure}

Figure~\ref{FIG:J1J2VNE} presents the variation of the entanglement entropy $\mathcal{S}_{vN}$, computed using the iDMRG method.
A pronounced enhancement of $\mathcal{S}_{vN}$ is observed near $J_{2,t}^{\textrm{var}}/J_1$, followed by a sharp drop across this region.
Notably, $\mathcal{S}_{vN}$ remains nearly constant as the bond dimension $m$ is increased from 1600 to 6400, suggesting that the ground state stays gapped even at the transition point. The inset displays the entanglement entropy within the narrow interval $0.760 \leq J_2/J_1 \leq 0.761$. The observed discontinuity in $\mathcal{S}_{vN}$ yields a transition point of $J_{2,t}/J_1 = 0.760616(2)$, thereby providing a definitive numerical benchmark that resolves longstanding uncertainty in earlier estimates.

To complement the above entanglement-based analysis, we now invoke the nonlocal SOPs to probe the topological character of this QPT.
The Haldane phase has the usual SOP $\mathcal{O}_1^z = \lim_{|j-i|\to\infty}\mathcal{O}_1^z(i,j)$ \cite{denNijs1989PRB}, where
\begin{equation}\label{EQ:SOP1}
\mathcal{O}_1^z(i,j) = -\left< S_i^z \big(e^{\imath\pi \sum_{k=i+1}^{j-1} S_k^z}\big) S_j^z \right>.
\end{equation}
In contrast, since the NNN-Haldane phase can be treated as two intertwined valence bond strings \cite{Kolezhuk2002PRB}, it is characterized by an alternative SOP $\mathcal{O}_2^z = \lim_{|j-i|\to\infty}\mathcal{O}_2^z(i,j)$ in which
\begin{equation}\label{EQ:SOP2}
\mathcal{O}_2^z(i,j) = -\left< S_i^zS_{i+1}^z \big(e^{\imath\pi \sum_{k=i+2}^{j-2} S_k^z}\big) S_{j-1}^zS_j^z \right>.
\end{equation}
One readily finds that $\mathcal{O}_2^z$ vanishes in the Haldane phase since $\mathcal{O}_2^z(i,j)$ can be factorized into $\langle S_i^zS_j^z\rangle \cdot \mathcal{O}_1^z(i+1,j-1)$.
The SOPs are numerically fragile and exhibit strong finite-size effects \cite{Anfuso2007}. To mitigate disturbances exacerbated by edge states, we perform large-scale DMRG calculations under PBC, retaining up to 4000 states. As shown in Fig.~\ref{FIG:J1J2SOP}(a), both types of SOPs display distinct jumps, with the discontinuities becoming more pronounced as the chain length $L$ increases. From Figs.~\ref{FIG:J1J2SOP}(b) and (c), it can be seen that their respective values drop from 0.0695 and 0.0562 to nearly zero for $\mathcal{O}_1^z$ and $\mathcal{O}_2^z$ at $L=160$.
This behavior is further corroborated by a corresponding jump in the bond-strength difference (see Supporting Information \cite{SuppMat}).

\begin{figure}[!t]
\centering
\includegraphics[width=0.90\columnwidth, clip]{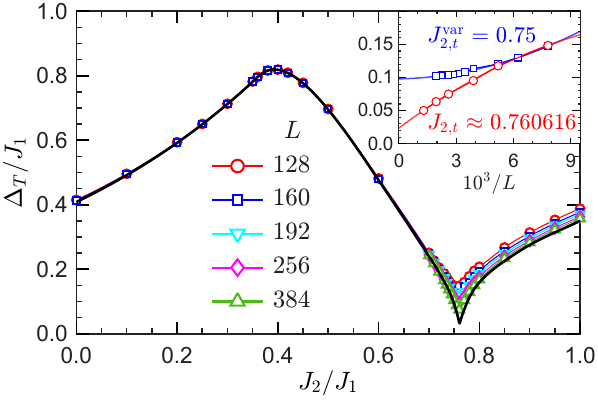}\\
  \caption{Unified excitation gap $\Delta_T$ versus $J_2/J_1$ for the $J_1$-$J_2$ chain. Data are shown for chain lengths $L$ ranging from 128 to 384. The thick solid line represents the $L \rightarrow \infty$ extrapolation of the gap. The inset illustrates the finite-size extrapolation procedure at two representative points: $J_{2,t}^{\text{var}} = 0.75$ (blue square) and $J_{2,t} = 0.760616(2)$ (red circle).}
  \label{FIG:J1J2GAP}
\end{figure}

To further clarify the nature of the QPT, we calculate the excitation gaps \cite{Kolezhuk1996PRL,Pati1996EPL}.
The Haldane phase has two spin-$1/2$ edge states for OBC, which account for
nearly degenerate ground states ($\textrm{S}_{\rm tot}\!=\!0,1$) with a
magnon excitation gap to higher spin states. The NNN-Haldane phase contains
effectively two chains, which couple to a unique ground state
($\textrm{S}_{\rm tot}\!=\!0$) without topological edge states. In this
case, each magnon excitation increases both the spin state and the energy by a
finite gap value. Hence, in both  phases the magnon excitation
gap is given by $\Delta_T = E_0(\textrm{S}_{\rm tot}\! =\! 2) -
E_0(\textrm{S}_{\rm tot} \!= \! 1)$. 
The transition is weakly first order, so the gap is strongly suppressed at the phase boundary but it does not vanish in the thermodynamic limit. 
As shown in Fig.~\ref{FIG:J1J2GAP}, $\Delta_T$ initially rises from $\Delta_H = 0.41050(2)$, signaling a stabilization of the Haldane phase under weak frustration. It then decreases from its peak, reaching a minimum at the transition point, after which it climbs steadily toward a saturation value.
For $J_2\!> \!J_1$, the edge
spins of the two chains in the NNN-Haldane phase are only weakly coupled by $J_1$, which determines the lowest-energy gap $\Delta_T \!\simeq\! 0.71 J_1$ for $J_2\!\gg\! J_1$ (not shown).
The inset of Fig.~\ref{FIG:J1J2GAP} reveals that at the transition point $J_{2,t}/J_1 = 0.760616(2)$, the excitation gap is about $0.023$, roughly a quarter of the value at $J_{2,t}^{\textrm{var}}/J_1 = 3/4$.
To conclude, the discontinuity in entanglement entropy, the abrupt jumps in nonlocal SOPs, and the finite excitation gap at the transition point collectively demonstrate that the Haldane--NNN-Haldane transition is of first order.
Notably, Vanderstraeten \textit{et al.} proposed recently that
the spinons could emerge in such a first-order transition \cite{Vanderstraeten2020}.
Since the spinons account for the formation and destruction of the NN and NNN valence bonds,
we speculate that these jump characters may relate to the abrupt confinement of spinons.

\begin{figure}[!t]
\centering
  \includegraphics[width=0.90\columnwidth, clip]{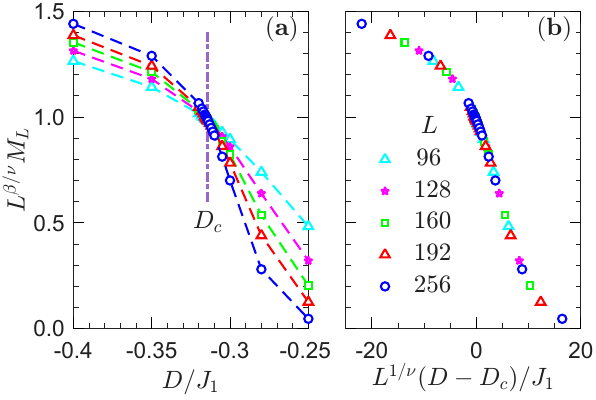}\\
  \caption{The FSS of the N\'eel order parameter $M_L$ versus $D/J_1$ for the Heisenberg chain at $J_2/J_1 = 0$. Data are shown for chain lengths $L = 96$ (red circles), $128$ (green triangles), $160$ (blue squares), $192$ (cyan pentagrams), and $256$ (pink diamonds).}
    \label{FIG:J2ZeroFSSNeel}
\end{figure}

\subsection{Haldane--N{\'e}el Ising transition}

We now examine the influence of a SIA $D$ on the Heisenberg model.
A negative $D$ introduces an easy-axis anisotropy along the $z$-direction, suppressing quantum spin fluctuations by energetically favoring states with maximal spin polarization along this axis. This suppression allows the intrinsic antiferromagnetic tendency of the Heisenberg interaction to develop into true long-range N{\'e}el order, characterized by a finite staggered magnetization.
The N{\'e}el phase is gapped and exhibits a two-fold degenerate ground state irrespective of the boundary condition.
To select one of the two degenerate ground states and thus define a meaningful order parameter $M_L^z$,
we apply small oppositely directed symmetry-breaking fields at the two edge spins. The local staggered magnetization in the central region remains robust after an initial decay near the edges, confirming its suitability as an order parameter.

The critical point and the corresponding critical exponents can be extracted via FSS.
Specifically, the order parameter $M_{L}^{z}$ obeys the scaling form
\begin{equation}\label{EQ:FSS}
M_L^{z} \simeq L^{-\beta_M/\nu}f_M\left(\vert D-D_c\vert L^{1/\nu}\right),
\end{equation}
where the critical exponent $\nu$ governs the divergence of the correlation length
and $\beta_M$ describes the scaling of the order parameter, $M\sim\vert D-D_c\vert^{\beta_M}$.
The exponents are determined by varying trial exponents $\mu_{1,2}$ until the rescaled quantity $M^{z}_{L}L^{\mu_{1}}$ collapses onto a universal function of $\vert D-D_{c}\vert L^{\mu_2}$ across all chain lengths $L$. This yields $\beta_M = \mu_1/\mu_2$ and $\nu = 1/\mu_2$.
Figure~\ref{FIG:J2ZeroFSSNeel} presents the FSS analysis of $M_L^{z}$.
The critical point is estimated as $D_{c} = -0.31448(4)$, in agreement with earlier studies \cite{Golinelli_1992,Albuquerque_2009,Chiara2012PRL,Lambert2019PRB}.
The extracted critical exponents, $\beta\simeq0.129$ and $\nu\simeq1.0$, are consistent with $\beta = 1/8$ and $\nu = 1$ for the two-dimensional Ising universality class.

\begin{figure}[!t]
  \centering
  \includegraphics[width=0.90\columnwidth, clip]{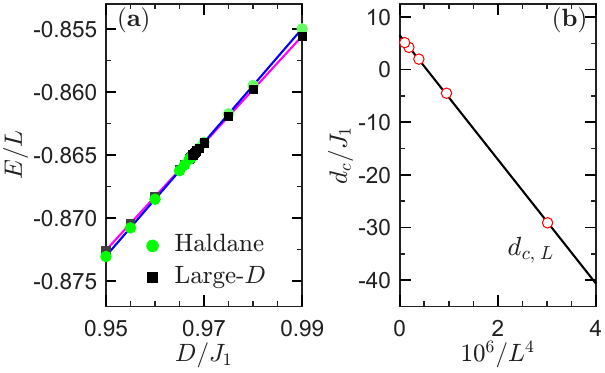}\\
  \caption{(a) Level crossing between the two lowest energy levels for $L = 32$ under TBC. Data points correspond to the Haldane phase (green circles) and the large-$D$ phase (black squares). (b) Richardson extrapolation of the critical point $D_c$ using the finite-size estimates $D_{c,L}$. Here, $D_{c,L}$ is defined as $0.96847250 + 10^{-8} d_{c,L}$, where $d_{c,L}$ is the last significant digit of $D_{c,L}$.}
  \label{FIG:J2ZeroDzCrit}
\end{figure}

\subsection{Haldane--large-$D$ topological phase transition}

Unlike the case of negative SIA $D$, which stabilizes magnetic order by imposing easy-axis alignment, a positive $D$ energetically selects the non-magnetic $S^z\! =\! 0$ state. This enhances quantum fluctuations, suppresses long-range magnetic order, and leads to a gapped paramagnetic regime known as the large-$D$ phase.
The transition between the Haldane phase and the large-$D$ phase represents one of the canonical examples of a continuous topological QPT. It belongs to the Gaussian universality class, described by a conformal field theory of a free boson with central charge $c = 1$ \cite{Hu2011PRB}.

Accurate determination of the Gaussian critical point $D_c$ is highly challenging and has attracted sustained interest, reflecting evolving theoretical perspectives over time \cite{Hu2011PRB,Tzeng2017PRB}.
Early studies relied on extrapolating peaks of the fidelity susceptibility or entanglement entropy computed via DMRG \cite{Tzeng_2008_1,Tzeng_2008_2}, which exhibit only linear convergence in $1/L$, typically limiting precision to two or three decimal places unless exploiting extremely large system sizes~\cite{Hu2011PRB}. The level spectroscopy method, enabled by the distinct parity and time-reversal eigenvalues under TBC~\cite{Tzeng2017PRB}, marked a major advance, offering quadratic $1/L^2$ convergence and establishing itself as a highly effective tool for analyzing the Haldane--large-$D$ transition~\cite{Kitazawa1996,Kitazawa1997a,Chen2003PRB,Ejima2015,Tzeng2017PRB}.
In this work, we build upon the level spectroscopy framework and achieve a further improvement in convergence. For each even chain length $L$, the level-crossing point $D_{c,L}$ obeys the scaling form
$D_{c,L}$ = $D_{c}$ + $c_2L^{-2}$ + $c_4L^{-4}$ + $\mathcal{O}(L^{-6})$
as established in Refs.~\cite{Kitazawa1996,Kitazawa1997a}.
By applying a Richardson-type extrapolation scheme, we elevate the convergence rate to quartic order in $1/L$.
Our numerical data are organized into two sets: $S_1 = \{L|24, 32, 40, 48, 56\}$ and $S_2 = \{L|48, 64, 80, 96, 112\}$. The critical point is extracted via the refined fitting formula
\begin{equation}\label{EQ:RenewedFit}
\overline{D}_{c,L} \equiv \frac{4D_{c,2L}-D_{c,L}}{3} \simeq D_{c} - \frac{c_4}{4L^{4}},
\end{equation}
which effectively suppresses the subleading corrections and isolates the dominant quartic convergence.
To ensure the reliability of Eq.~\eqref{EQ:RenewedFit}, we retain up to 6000 block states and perform up to 24 sweeps, maintaining the energy level spacing precision to at least $10^{-8}$.

\begin{table}[th!]
\caption{\label{TAB:J2ZeroLS}
    The level-crossing quasi-critical points $D_{c,L}$'s at different lengths.}
    \begin{ruledtabular}
    \begin{tabular}{ l c | l c | l c}
    $L$ & $D_{c,L}$ & $L$ & $D_{c,L}$ & $L$ & $D_{c,L}$\\
    \colrule
    24   & 0.9671760094\;  & 32   & 0.9677428936\;  & 40   & 0.9680054750\\
    48   & 0.9681481590\;  & 56   & 0.9682342060\;  & 64   & 0.9682900648\\
    80   & 0.9683557587\;  & 96   & 0.9683914464\;  & 112  & 0.9684129651
    \end{tabular}
    \end{ruledtabular}
\end{table}

Figure~\ref{FIG:J2ZeroDzCrit}(a) displays the two lowest energy levels corresponding to the Haldane and large-$D$ phases for $L = 32$. The level crossing point $D_{c,L}$ is clearly identified and determined with high precision via a stepwise predictor-corrector procedure, yielding accuracy beyond eight decimal places. The finite-size quasi-critical points $D_{c,L}$ for chain lengths $L$ = 24, 32, 40, 48, 56, 64, 80, 96, and 112 are summarized in Table~\ref{TAB:J2ZeroLS}.
Fitting the data to the scaling formula of $D_{c,L}$ with either linear or quadratic terms in $1/L^2$ gives critical points of $0.9684724(3)$ and $0.9684725(3)$, respectively. Alternatively, applying the refined expression from Eq.~\eqref{EQ:RenewedFit} allows us to separate the leading estimate 0.96847250 from a residual term of order $10^{-8} d_{c,L}$, where $d_{c,L}$ represents the eighth decimal digit of $D_c$. As shown in Fig.~\ref{FIG:J2ZeroDzCrit}(b), we obtain $d_{c,L} = 6(2)$. Consequently, the Gaussian critical point is determined to be $D_c = 0.96847256(2)$.

\section{Spontaneous dimerization and deconfined criticality}\label{SEC:Dimer}

\subsection{Perturbation analysis of the dimerized phase}

For the spin-1 $J_1$--$J_2$ chain with strong easy-axis SIA $D \ll 0$, the low-energy sector is restricted to the $S^z = \pm 1$ doublet. 
Using second-order perturbation theory, we recast this model into an effective pseudospin-$1/2$ Hamiltonian
\begin{align}\label{EQ:PertJ1J2Ising}
H_{\text{eff}} \simeq   & \sum_i (J_1 S_i^z S_{i+1}^z + J_2 S_i^z S_{i+2}^z) \nonumber\\
                        & - \sum_i \sum_{\delta = 1,2} \frac{J_{\delta}^2}{|D|} \left( S_i^x S_{i+\delta}^x + S_i^y S_{i+\delta}^y \right),
\end{align}
where $\mathbf{S}_i$ are spin-$1/2$ operators within the projected subspace. In the limit $|D| \to \infty$, the model reduces to the classical $J_1$--$J_2$ Ising chain, which exhibits a first-order transition between the N{\'e}el phase ($\uparrow\downarrow$) and the NNN-N{\'e}el phase ($\uparrow\uparrow\downarrow\downarrow$) at $J_2/J_1 = 1/2$ \cite{Igarashi1989PRB,Heidrich-Meisner_2007}.
The gapped NNN-N{\'e}el phase has a two~(four)-fold ground-state degeneracy under OBC~(PBC). Under OBC, the corresponding order parameter is extracted by applying a uniform pinning field to the spins at both ends of the chain.
Physically, the NNN-N{\'e}el phase can also be realized by other types of exchange interactions \cite{Harada1993JPSJ,Kaplan2009PRB}, and is argued to be the ground state of a series of spinel compounds \cite{Zou2016PRB}.

In contrast, a finite $|D|$ introduces transverse exchange terms $\propto J_{\delta}^2/|D|$, which generate quantum fluctuations.
These fluctuations play a decisive role near the classical critical point. They lift the degeneracy between competing classical orders via second-order energy corrections that favor local singlet formation. A translationally invariant chain cannot fully optimize this energy gain; instead, a spontaneously dimerized pattern emerges, characterized by strong valence bonds on alternating $J_1$ links. This configuration maximizes the energy lowering from transverse exchange through enhanced local resonance, thereby stabilizing a gapped dimerized phase via an order-by-disorder mechanism. Consequently, the classical first-order transition is superseded by an extended quantum dimerized phase, bounded by continuous transition lines, in direct analogy to the frustrated transverse-field Ising chain.

\begin{figure}[!ht]
  \centering
  \includegraphics[width=0.95\columnwidth, clip]{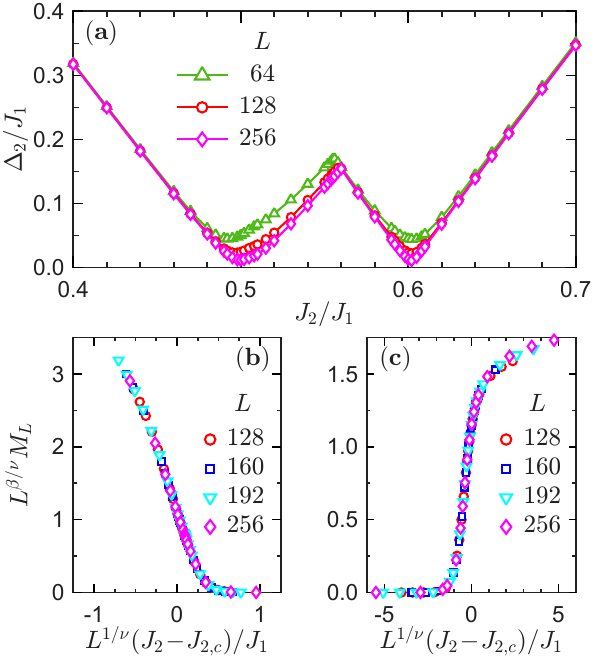}\\
  \caption{(a) Excitation gap $\Delta_2$ as a function of $J_2/J_1$ at $D/J_1 = -2$ for chain lengths $L = 64$ (circles), $128$ (triangles), and $256$ (squares). The gap minimum identifies the transition point.
  Finite-size scaling of the staggered magnetization $M_L^z$ across the (b) N{\'e}el--dimer and (c) NNN-N{\'e}el--dimer transitions along the same cut.}
  \label{FIG:DzM200GAPFSS}
\end{figure}

\subsection{Deconfined quantum critical point}

A symmetry analysis of the competing phases is essential for understanding the nature of their QPTs and the appropriate theoretical framework. The N\'eel phase breaks the discrete spin-inversion symmetry $\mathbb{Z}_2^{S}$ while preserving translation, whereas both the NNN-N\'eel phase and the dimerized phase break translational symmetry $\mathbb{Z}_2^{T}$ but preserve $\mathbb{Z}_2^{S}$. This distinction in symmetry breaking directly governs the universality of the QPTs. A transition between the NNN-N\'eel and dimerized phases, which break the same $\mathbb{Z}_2^{T}$ symmetry, can be described within the LGW paradigm by a single real order parameter and is expected to belong to the Ising universality class. In contrast, a transition between the N\'eel and dimerized phases involves states that break different, non-commuting symmetries ($\mathbb{Z}_2^{S}$ vs. $\mathbb{Z}_2^{T}$). According to LGW theory, such a transition is generically first-order; a direct continuous QPT between them therefore lies outside the LGW framework and may require a description in terms of deconfined fractionalized excitations or a non-trivial conformal field theory \cite{Jiang2019PRB,Roberts2019PRB,Mudry2019PRB}. This symmetry-based dichotomy fundamentally explains the distinct critical behaviors observed in the phase diagram of the frustrated anisotropic spin chain.

To elucidate the nature of the QPTs, we carry out a systematic analysis along the line defined by $D/J_1 = -2$. Under OBC, the ground states of the N\'eel and NNN-N\'eel phases are two-fold degenerate, whereas the intermediate dimerized phase exhibits a unique ground state. This distinction allows us to employ the second excitation gap, $\Delta_2 = E_2 - E_0$, to discriminate between the different phases. As shown in Fig.~\ref{FIG:DzM200GAPFSS}(a), the calculated energy gap displays two pronounced minima near $ J_{2,c}^{(1)}/J_1 \approx 0.5$ and $ J_{2,c}^{(2)}/J_1 \approx 0.6$. The fact that these dips approach zero identifies them as signatures of continuous QPTs, which separate the N\'eel, dimer, and NNN-N\'eel phases.

\begin{figure}[!ht]
  \centering
  \includegraphics[width=0.95\columnwidth, clip]{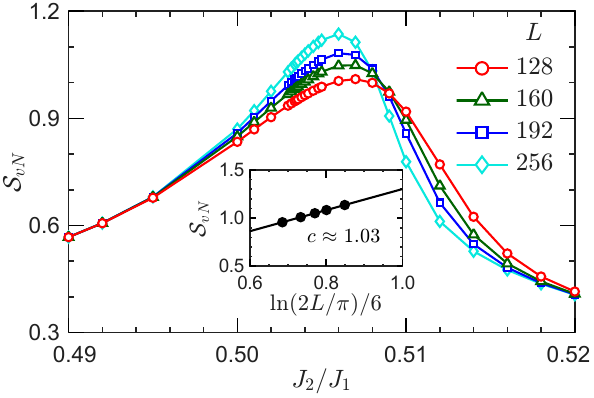}\\
  \caption{Entanglement entropy $\mathcal{S}_{vN}$ versus $J_2/J_1$ for the $J_1$-$J_2$-$D$ chain at $D/J_1 = -2$, shown for chain length $L = 128$ (red circles), $160$ (green triangles), $192$ (blue squares), and $256$ (cyan diamonds). The inset shows the finite-size scaling of the peak values $\mathcal{S}_{vN}^{\text{max}}(L)$. The solid line is a fit to Eq.~\eqref{EQ:VNE}, from which the central charge $c$ is extracted.}
  \label{FIG:DzM200CFT}
\end{figure}

To quantify the critical behavior around these transition points, we conduct a FSS analysis of the order parameters. The scaling collapses for the magnetically ordered N\'eel and NNN-N\'eel phases are presented in Figs.~\ref{FIG:DzM200GAPFSS}(b) and (c), respectively. For the transition between the NNN-N\'eel and dimer phases, the extracted critical exponents $\beta = 0.127(4)$ and $\nu = 1.03(5)$ are consistent with the two-dimensional Ising universality class. In contrast, for the N\'eel--dimer transition, the obtained exponents $\beta = 0.44(2)$ and $\nu = 1.35(3)$ do not align with any conventional universality class. Notably, they approximately satisfy the scaling relation $2(\nu - 2\beta) = 1$, a hallmark of the eight-vertex universality class in certain deconfined critical scenarios \cite{Jiang2019PRB,Roberts2019PRB,Pronk2025PRB}. To further confirm this identification, we compute the central charge $c$ via entanglement entropy scaling. Figure~\ref{FIG:DzM200CFT} shows the evolution of the entanglement entropy across the critical point at $J_{2,c}^{(1)}/J_1 = 0.50432(1)$, where a distinct peak emerges. As the chain length $L$ increases, this peak sharpens and its maximum value grows. By fitting the peak maxima using Eq.~\eqref{EQ:VNE} (see the inset), we extract a central charge $c = 1.03(4)$, a value close to unity, consistent with the free-boson description expected for the eight-vertex universality class.

\begin{table}[!b]
\vspace{-0.50cm}
\caption{\label{TAB:GauFSS}
Extracted critical point $J_{2,c}$ and critical exponents $\beta$ and $\nu$ of the continuous N{\'e}el--dimer phase transition.}
\begin{ruledtabular}
\begin{tabular}{ c | c c c c c c}
$D/J_1$  & $J_{2,c}/J_1$ & $\beta/\nu$ & $1/\nu$ & $\beta$ & $\nu$ & $2(\nu-2\beta)$  \\
\colrule
-5.00    & 0.50511  & 0.3720 & 0.54 & 0.69   & 1.85  & 0.94\\
-4.50    & 0.50531  & 0.3705 & 0.56 & 0.66   & 1.80  & 0.93\\
-4.00    & 0.50545  & 0.3672 & 0.57 & 0.64   & 1.75  & 0.93\\
-3.50    & 0.50550  & 0.3625 & 0.59 & 0.62   & 1.70  & 0.93\\
-3.00    & 0.50538  & 0.3558 & 0.62 & 0.57   & 1.60  & 0.92\\
-2.50    & 0.50501  & 0.3467 & 0.65 & 0.54   & 1.55  & 0.95\\
-2.00    & 0.50432  & 0.3285 & 0.74 & 0.44   & 1.35  & 0.93\\
\end{tabular}
\end{ruledtabular}
\end{table}

Moving beyond the specific cut at $D/J_1 = -2$, our analysis confirms that the N\'eel--dimer transition retains its eight-vertex character across a broader parameter range. The complete set of critical points and associated exponents for various SIA strengths is summarized in Table~\ref{TAB:GauFSS}. Analysis of the critical points reveals that the phase boundary as a function of $D$ is non-monotonic, exhibiting a slight hump near $D \approx -3.50$ (see Supporting Information \cite{SuppMat}). Furthermore, while the critical exponents $\beta$ and $\nu$ vary with the SIA, they consistently and approximately satisfy $2(\nu - 2\beta) = 1$ across the entire boundary, reinforcing the eight-vertex nature of this DQCP.

\begin{figure}[t]
  \centering
  \includegraphics[width=0.90\columnwidth, clip]{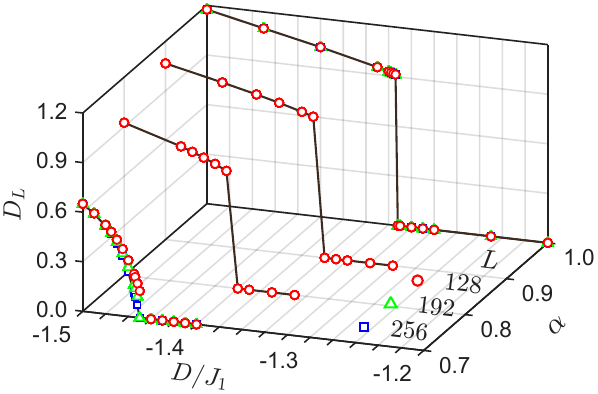}\\
  \caption{The dimer order parameter $\mathcal{D}_L$ versus $D/J_1$ for the $J_1$-$J_2$-$D$ chain along the cut $J_2$ = 7/30 ($\alpha - D$) with $\alpha$ = 0.7, 0.8, 0.9, and 1.0. Data are shown for chain lengths $L = 128$ (red circles), $192$ (green triangles), and $256$ (blue squares).}
  \label{FIG:DimerHDOP}
\end{figure}

\subsection{Absence of changeover phenomenon}

We now turn to the transitions emanating from the dimerized phase into the disordered regimes. The dimer order is captured by the staggered correlator
\begin{equation}
D_L = \langle S_{L/2}^{z}S_{L/2-1}^{z} - S_{L/2}^{z}S_{L/2+1}^{z}\rangle,
\end{equation}
which measures the local imbalance of nearest-neighbor spin correlations across a bond.

Across the transition between the dimerized and NNN-Haldane phases, $D_L$ varies continuously, and its FSS behavior unambiguously places the transition in the Ising universality class (see Supporting Information \cite{SuppMat}). We next examine the dimer--Haldane boundary. Figure~\ref{FIG:DimerHDOP} presents $D_L$ along a family of parallel cuts parametrized by $J_2 = (7/30)(\alpha - D)$, with $\alpha$ decreasing from $1.0$ to $0.7$, approaching the N\'eel--Haldane--dimer triple point. At each cut, $D_L$ exhibits a clear discontinuity at the transition. This jump, though somewhat suppressed near $\alpha = 0.7$ owing to the proximity of the triple point, becomes increasingly pronounced with growing $\alpha$, consistently attesting to a first-order dimer--Haldane transition. This scenario is further corroborated by the concomitant discontinuities observed in the entanglement entropy across the transition.

The nature of the dimer--Haldane transition has attracted sustained attention in the study of frustrated spin-1 chains. Depending on the microscopic parameters, this transition has been reported to realize diverse universality classes, ranging from first-order to $SU(2)_{2}$ Wess-Zumino-Witten criticality \cite{Chepiga_2016_1,Chepiga_2016_2,Luo2026arXiv}. While such a continuous scenario cannot be definitively ruled out by numerical means alone, our data offer no indication of this possibility in the present model. Importantly, we find no evidence of a changeover from first-order to continuous behavior upon approaching the triple point --- a phenomenon observed in certain other frustrated spin systems \cite{Chepiga_2016_1,Chepiga_2016_2,Luo2026arXiv}. The absence of this crossover in our model underscores the subtle interplay between frustration and anisotropy, and calls for a refined theoretical understanding of the dimer--Haldane transition.

\section{Emergent chirality and weakly first-order transition}\label{SEC:Chiral}

In strongly frustrated spin chains with easy-plane anisotropy, a moderate $D>0$ selectively suppresses out-of-plane fluctuations while preserving the chiral degrees of freedom associated with the spiral instability, thereby stabilizing a phase with spontaneously broken inversion symmetry, signaled by a finite vector spin chirality \cite{Okunishi2008,He2024PRB}
\begin{equation}
\kappa_n^z = \langle (\mathbf{S}_n \times \mathbf{S}_{n+1})^z \rangle = -\frac{i}{2}\langle S_n^+ S_{n+1}^- - S_n^- S_{n+1}^+ \rangle.
\end{equation}

\subsection{Bosonization analysis of the chiral phase}

The emergence of the chiral phase can be understood within a bosonization framework in the limit of small frustrating coupling $J_1$ between two chains.  The transition from Haldane to large-$D$ phase near $D_c$ for non-frustrated chains ($J_1\!\to\! 0$) has been predicted using bosonization quite a while ago \cite{Schulz1986}, which we shortly review following the notation in Ref.~\cite{Berg2008}, where we replace $U\to 2D$ and $V=2t\to J_2$:
Each spin-1 operator is represented as a sum of two spin-1/2 fields, with the Hilbert space restricted to the triplet sector.
The Hamiltonian can be expressed in terms of symmetric (+) and anti-symmetric ($-$) combinations of those fields \cite{Berg2008}:
\begin{subequations}\label{EQ:HcHs}
\begin{align}
\mathcal{H}_+ = & \frac{v_+}{2\pi}\left[K_+(\partial_x\theta_+)^2+\frac{1}{K_+}(\partial_x\phi_+)^2\right] \\ \nonumber
                & + g_1\cos(2\phi_+), \\
\mathcal{H}_- = & \frac{v_-}{2\pi}\left[K_-(\partial_x\theta_-)^2+\frac{1}{K_-}(\partial_x\phi_-)^2\right] \\ \nonumber
                & + g_2\cos(2\phi_-) + g_3\cos(2\theta_-),
\end{align}
\end{subequations}
where the irrelevant inter-sector couplings have been omitted. The coefficients in Eq.~\eqref{EQ:HcHs} are determined as \cite{Berg2008}
$K_+ \! =\! 2[1+4(D/J_2+3)/\pi]^{-1/2}$, $K_- \!=\! [1+4(1-D/J_2)/\pi]^{-1/2}$, $g_1\! =\! -g_2\! \propto\! J_2\!-\!D$, and $g_3 \!\propto\! -J_2$.
Near $D_c\sim J_2$, both $g_3$ and $g_1$ are relevant with scaling dimensions $1/K_-$ and $K_+<2$, respectively.  The transition is driven by a sign change 
of $g_1$ in the symmetric sector, leading to a $c\!=\!1$ phase boundary between  Haldane and large-$D$ insulators.

To study the emergence of the chiral phase for small $J_1$ near the critical point, we use two sets of fields $a$ and $b$, one for each chain, which are coupled by $J_1$.
The antisymmetric fields remain frozen and the most relevant coupling terms in the symmetric sector are given by 
\begin{equation} \label{hi}
\mathcal{H}_{I}\! =\! 2 J_1 \partial_x\phi_+^a\partial_x\phi_+^b/\pi^2\! + \!\tilde g\,\partial_x(\theta_+^a\!-\!\theta_+^b)\sin\left({\theta_+^a\!-\!\theta_+^b}\right) , 
\end{equation}
where $\tilde g \propto J_1$.  Here, the first term stems from the uniform density-density coupling, which is exactly marginal and can be absorbed by rescaling \cite{Berg2008}. The second coupling constant $\tilde g$ with scaling dimension $1\!+\!1/(2\tilde K_+)$ comes from coupling the alternating spin-spin terms and is relevant for $\tilde K_+>1/2$. In contrast to Refs.~\cite{Berg2008,Giamarchi2001} where  non-frustrated spin ladders were considered, our frustrated geometry leads to an overall derivative of the corresponding coupling in Eq.~\eqref{hi}. 
Near the critical point we can assume the numerically determined value of $\tilde K_+\!\approx\! K_+\!\approx\! 1.321$ \cite{Hu2011PRB} since the 
rescaling from the marginal operator only leads to higher order corrections in $J_1$.   The competition of two relevant couplings $g_1$ and $\tilde g$ in renormalization group equations gives characteristic power-law shapes of the phase boundaries \cite{Berg2008,Giamarchi2001,MetavEggert2017}
\begin{equation}
|D-D_c| \propto J_1^{\alpha},\quad \alpha = \frac{2-K_+}{1-1/(2K_+)}\approx 1.09,
\end{equation}
consistent with the observed shape in Fig.~\ref{FIG:GSPD}(a). In particular, if $|D-D_c|$ is below this threshold powerlaw, the coupling $\tilde g$ dominates, leading to the emergence of a different order parameter. In Refs.~\cite{MetavEggert2017,Nersesyan1998} it has been argued that the corresponding inter-chain derivative operator drives incommensurate $S^+S^-$ spin-nematic correlations -- a phase identified as the chiral phase here.

\begin{figure}[!t]
  \centering
  \includegraphics[width=0.90\columnwidth, clip]{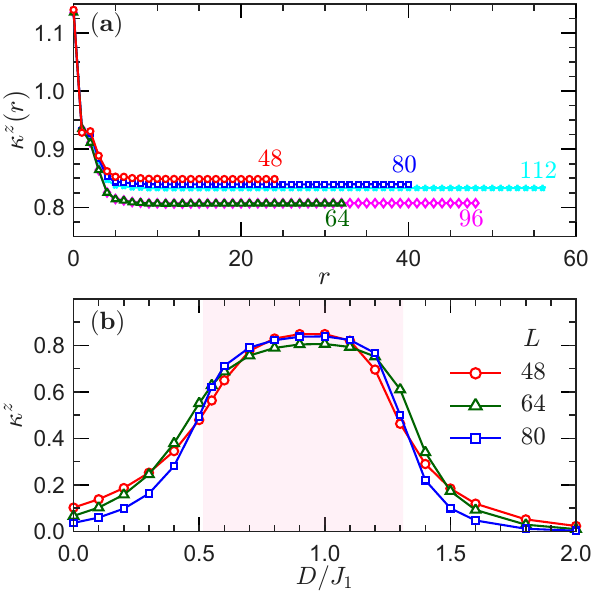}\\
  \caption{(a) The extracted chiral order parameter $\kappa(r)$ via the chiral-chiral correlation function at $D/J_1 = 1$ and $J_2/J_1 = 1$. Data are shown for chain length ranging from 48 to 112 under PBC. (b) The chiral order parameter $\kappa_L$ versus $D/J_1$ for the $J_1$-$J_2$-$D$ chain along the cut $J_2/J_1 = 1$, shown for chain lengths $L = 48$ (red circles), $64$ (green triangles), and $80$ (blue squares).}
    \label{FIG:J2100Chiral}
\end{figure}

\subsection{Chiral phase boundaries and critical properties}

We now turn to a detailed numerical characterization of the chiral phase at $J_2/J_1 = 1$, where the bosonization analysis no longer applies. To detect spontaneous chiral order, we note that the Hamiltonian is real-symmetric and admits a real ground state, whereas the chirality operator $\hat{\kappa}_n^z$ is purely imaginary --- hence $\langle \hat{\kappa}_n^z \rangle$ vanishes in any real eigenstate. We therefore extract the order parameter from the asymptotic behavior of the chiral-chiral correlation function:
\begin{equation}\label{EQ:ChiChi}
|\langle \hat{\kappa}^z \rangle|^2 = \lim_{|j-i| \gg 1} \langle \hat{\kappa}_i^z \hat{\kappa}_j^z \rangle,
\end{equation}
which allows us to determine $\langle \hat{\kappa}^z \rangle$ within the original real-symmetric representation \cite{Okunishi2008,He2024PRB}.

Figure~\ref{FIG:J2100Chiral} displays the chiral order parameter along $J_2/J_1 = 1$. Figure~\ref{FIG:J2100Chiral}(a) shows the correlation function for various chain lengths $L$ under PBC at $D/J_1 = 1$; the correlators saturate to a finite plateau, from which we extract the finite-size order parameter $\langle \hat{\kappa}_L^z \rangle$ by evaluating the correlator between sites $i=1$ and $j=L/2$. Despite nonmonotonic finite-size variations, the extracted chiral order exhibits clear convergence to a finite value in the thermodynamic limit. Figure~\ref{FIG:J2100Chiral}(b) summarizes $\langle \hat{\kappa}_L^z \rangle$ across $0 \leq D/J_1 \leq 2$; the order parameter remains finite within $0.5 \lesssim D/J_1 \lesssim 1.3$, suggesting continuous QPTs to the adjacent NNN-Haldane and large-$D$ phases.
\begin{figure}[!t]
  \centering
  \includegraphics[width=0.95\columnwidth, clip]{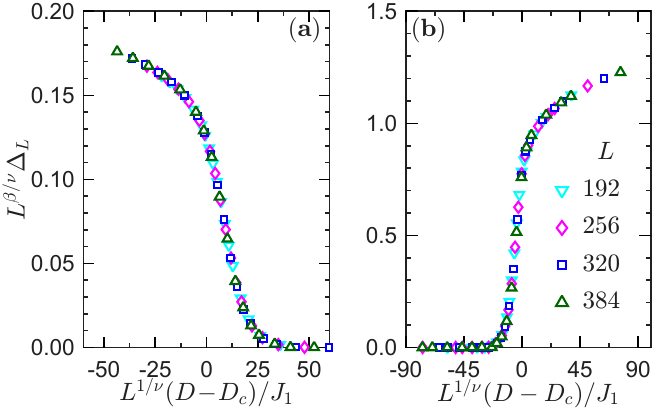}\\
  \caption{The FSS of the Schmidt gap $\Delta_L$ for (a) the NNN-Haldane--Chiral transition and (b) the Large-$D$--Chiral transition in the $J_1$-$J_2$-$D$ chain at $J_2/J_1 = 1$. Data are shown for chain length ranging from 192 to 384. The critical points and critical exponents are shown in the main text.}
\label{FIG:J2100SchmidtGapFSS}
\end{figure}

To pinpoint the phase boundaries, we employ the Schmidt gap $\Delta_L$, defined as the difference between the two largest eigenvalues of the reduced density matrix \cite{Li2008PRL}. The ground-state wave function of the chiral phase is real, originating from an equal-weight superposition of two degenerate complex wave functions, yielding a trivial twofold degeneracy in the entanglement spectrum. This exemplifies that an even-fold degeneracy is not a sufficient identifier of SPT phases --- indeed, the columnar dimer phase (with spin-1/2 edge states \cite{Luo2018}) and the N\'eel phase (with spontaneous $\mathbb{Z}_2$ symmetry breaking \cite{Tzeng2017PRB}) can also display twofold degeneracy. In contrast, the NNN-Haldane and large-$D$ phases possess non-degenerate entanglement spectra, making the Schmidt gap an effective tool for locating the boundaries. Figure~\ref{FIG:J2100SchmidtGapFSS} presents the FSS of $\Delta_L$ along $J_2/J_1 = 1.00$, yielding critical points at $D_{c,1}/J_1 = 0.5141(5)$ and $D_{c,2}/J_1 = 1.3101(3)$, with the extracted critical exponents consistent with the Ising universality class at both transitions.

\subsection{Weakly first-order transition between distinct chiral phases}

A distinct phase -- the Haldane-chiral phase -- emerges in a narrow region between the Haldane and ordinary chiral phases. Under PBC, it exhibits a twofold degeneracy from spontaneous $\mathbb{Z}_2$ chiral symmetry breaking; under OBC, this multiplies to eightfold due to the combination with the fourfold edge-state degeneracy of the Haldane phase. This stark contrast directly manifests the intertwined symmetry-breaking and topological orders of this phase.

\begin{figure}[!t]
  \centering
  \includegraphics[width=0.95\columnwidth, clip]{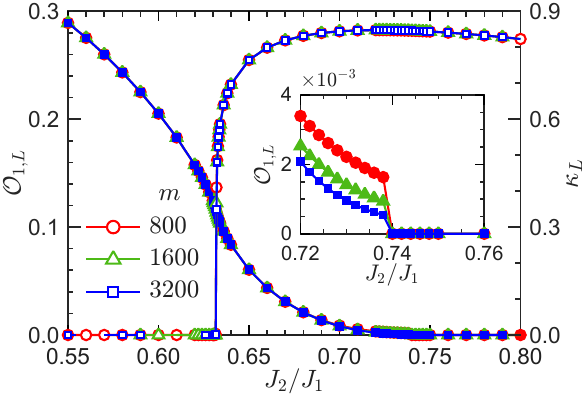}\\
  \caption{(a) The SOP $\mathcal{O}_{1,L}$ (left axis, filled symbols) and chiral order $\kappa_L$ (right axis, open symbols) in the $J_1$-$J_2$-$D$ chain at $D/J_1 = 0.5$. The truncated dimension $m$ kept in the iDMRG calculation is 800 (red circles), 1600 (green triangles), and 3200 (blue squares). The inset magnifies $\mathcal{O}_{1,L}$ near the Haldane-chiral--chiral transition in the region where $0.72 \leq J_2/J_1 \leq 0.76$.}
  \label{FIG:Dz050ChiSOP}
\end{figure}

We establish its existence by simultaneously computing the SOP $\mathcal{O}^z$ [Eq.~\eqref{EQ:SOP1}] and the chiral order parameter $\kappa^z$ [Eq.~\eqref{EQ:ChiChi}] using iDMRG on infinite systems. Figure~\ref{FIG:Dz050ChiSOP} presents $\mathcal{O}_{1,L}$ (filled symbols) and $\kappa_{L}$ (open symbols) as functions of $J_2/J_1$ at fixed $D/J_1 = 0.5$. As $J_2/J_1$ increases, $\mathcal{O}_{1,L}$ gradually decreases, shows a smooth downturn at $J_{2,t}^{(1)}/J_1 \!= \!0.6318(2)$, and then drops sharply to zero at $J_{2,t}^{(2)}/J_1 \!=\! 0.739(1)$ (see inset). Meanwhile, $\kappa_L$ emerges abruptly at $J_{2,t}^{(1)}/J_1$ and remains finite across the entire intervening region. The coexistence of both order parameters identifies the Haldane-chiral phase.

The transition at $J_{2,t}^{(1)}/J_1$ (Haldane to Haldane-chiral) is of Ising type. In contrast, the transition at $J_{2,t}^{(2)}/J_1$ (Haldane-chiral to ordinary chiral) is weakly first order: while $\kappa_L$ evolves continuously, $\mathcal{O}_{1,L}$ exhibits a subtle jump before vanishing, rendering the transition numerically difficult to resolve. This subtlety highlights the value of our high-precision diagnostics and provides a clear example of a weak first-order transition in a frustrated spin chain, directly addressing the central theme of this work.

\section{Conclusions}\label{SEC:CONC}
In summary, we have systematically mapped the ground-state phase diagram of a frustrated spin-1 chain with SIA, employing large-scale DMRG, iDMRG, and complementary analytical approaches. The model hosts eight distinct phases, with the most salient emergent ones being a spontaneous dimerized phase at negative SIA and two vector chiral phases at positive SIA. The dimerized phase arises from an order-by-disorder mechanism, while the vector chiral orders originate from a spin-sector coupling term -- identified via bosonization as an in-plane chiral operator -- that dominates over conventional current-current interactions in the frustrated regime.

We precisely pinpoint several phase boundaries and their critical characters. The transition between the dimerized and NNN-N\'eel phases is Ising-type, whereas the dimer--N\'eel transition is continuous with central charge $c=1$ and falls into the eight-vertex universality class, placing it outside the LGW paradigm. Within this class, the critical exponents $\beta$ and $\nu$ vary continuously along the phase boundary while satisfying the hyperscaling relation $2(\nu-2\beta)=1$, revealing a tunable non-Landau criticality with a robust underlying scaling structure. For the positive-anisotropy side, the conventional chiral phase borders the NNN-Haldane and large-$D$ phases via Ising transitions, as confirmed by finite-size scaling of the Schmidt gap. A narrow Haldane-chiral phase --- characterized by coexisting string and chiral orders --- intervenes between the Haldane and conventional chiral phases; its transition to the Haldane phase is Ising-type, while its transition to the conventional chiral phase is weakly first order, as evidenced by a soft jump in the string order parameter.

Two representative topological transitions are pinpointed with exceptional precision. The weakly first-order Haldane--NNN-Haldane transition at $J_{2,t}\!=\!0.760616(2)$ is identified via the discontinuity in the entanglement entropy, with its topological nature further corroborated by the nonlocal string order parameters and the nonzero excitation gap at the transition point. The continuous Haldane--large-$D$ transition at $D_c\!=\!0.96847256(2)$, resolved by Richardson-extrapolated level spectroscopy, establishes a high-precision reference for excitation spectra and quantum criticality.

Finally, our results raise an intriguing open question regarding the changeover from first-order to continuous QPT as system parameters vary. While such a changeover between dimerized and Haldane phases has been reported in other spin-1 chains, in our model the dimer--Haldane transition remains consistently first order across the entire accessible parameter range, indicating the absence of this crossover and inviting further theoretical investigation into its microscopic origin.


\begin{acknowledgements}
This work is supported by the National Program on Key Research Project (Grant No. MOST2022YFA1402700),
the National Natural Science Foundation of China (Grants Nos. 12674185, 12674178, 12574163, and 12274187),
the Deutsche Forschungsgemeinschaft (DFG, German Research 278 Foundation)-Project No. 277625399-TRR 185 OSCAR (A5),
and the Beijing National Laboratory for Condensed Matter Physics (Grant No. 2025BNLCMPKF022).
The computations are partially supported by High Performance Computing Platform of Nanjing University of Aeronautics and Astronautics.
\end{acknowledgements}

\section*{DATA AVAILABILITY}
The data that support the findings of this article are not publicly available. The data are available from the authors upon reasonable request.




%




\clearpage

\onecolumngrid

\newpage

\newcounter{sectionSM}
\newcounter{equationSM}
\newcounter{figureSM}
\newcounter{tableSM}
\stepcounter{equationSM}
\setcounter{section}{0}
\setcounter{equation}{0}
\setcounter{figure}{0}
\setcounter{table}{0}
\setcounter{page}{1}
\makeatletter
\renewcommand{\thesection}{\textsc{S}\arabic{section}}
\renewcommand{\theequation}{\textsc{S}\arabic{equation}}
\renewcommand{\thefigure}{\textsc{S}\arabic{figure}}
\renewcommand{\thetable}{\textsc{S}\arabic{table}}


\begin{center}
{\large{\bf Supplemental Material for\\
``Nontrivial criticality and weakly first-order transitions in the frustrated spin-1 chain\\ with single-ion anisotropy''}}
\end{center}
\begin{center}
Qiang Luo$^{1,\,2}$, Shijie Hu$^{3,\,4}$, Jize Zhao$^{5,\,6}$, Sebastian Eggert$^{4}$, and Xiaoqun Wang$^{7,\,8}$\\
\quad\\
$^1$\textit{College of Physics, Nanjing University of Aeronautics and Astronautics, Nanjing, 211106, China}\\
$^2$\textit{School of Physics, Renmin University of China, Beijing 100872, China}\\
$^3$\textit{Beijing Computational Science Research Center, Beijing 100084, China}\\
$^4$\textit{Physics Department and Research Center OPTIMAS, RPTU University Kaiserslautern-Landau, 67663 Kaiserslautern, Germany}\\
$^5$\textit{School of Physical Science and Technology $\&$ Key Laboratory of Quantum Theory and Applications of MoE, Lanzhou University, Lanzhou 730000, China}\\
$^6$\textit{Lanzhou Center for Theoretical Physics, Key Laboratory of Theoretical Physics of Gansu Province, Lanzhou University, Lanzhou 730000, China}\\
$^7$\textit{School of Physics and Institute for Advanced Study in Physics, Zhejiang University, Hangzhou 310058, China}\\
$^8$\textit{School of Physics and Astronomy, Tsung-Dao Lee Institute,\\ Shanghai Jiao Tong University, Shanghai 200240, China}\\

(Dated: Nov. 7th, 2026)
\quad\\
\end{center}

\onecolumngrid

This Supplemental Material presents results for variational derivation of the excitation dispersion, the bond-strength difference calculations, the schematic cuts emanating from the dimerized phase, and the dimer--NNN-Haldane transition analysis.

\vspace{0.50cm}



\section{Variational derivation of the excitation dispersion}
\label{sec:variational}

In this section we present a self-contained variational calculation of the elementary excitation spectrum of the Hamiltonian
\begin{equation}
\hat H=\sum_i \mathbf S_i\cdot \mathbf S_{i+1}
+\alpha\sum_i \mathbf S_i\cdot \mathbf S_{i+2}
\end{equation}
within the Affleck-Kennedy-Lieb-Tasaki (AKLT) matrix-product-state (MPS) framework. The derivation yields an analytic dispersion that reproduces the known result of Ref.~\cite{SMKolezhuk1997PRB} and provides a transparent physical picture of the first-order transition from the AKLT phase (i.e., Haldane phase) to the NNN-AKLT phase (i.e., NNN-Haldane phase). The section is organized as follows: Sec.~\ref{sec:AKLT_MPS} introduces the MPS representation and the soliton (crackion) states; Sec.~\ref{sec:norm_num} defines the excitation energy and computes the per-site norm and the per-site energy numerator; Sec.~\ref{sec:dispersion} presents the final dispersion and discusses its physical consequences.

\subsection{Matrix product states and soliton excitations}
\label{sec:AKLT_MPS}

We adopt the standard AKLT representation with matrices normalized so that the transfer matrix has maximal eigenvalue $1$,
\begin{equation}
C^+=
\begin{pmatrix}
0&a\\
0&0
\end{pmatrix},\qquad
C^0=
\begin{pmatrix}
-b&0\\
0&b
\end{pmatrix},\qquad
C^-=
\begin{pmatrix}
0&0\\
-a&0
\end{pmatrix},
\end{equation}
where $a=\sqrt{2/3}$ and $b=1/\sqrt{3}$. The ground-state MPS is
\begin{equation}
|\mathrm{AKLT}\rangle
=
\mathrm{Tr}\left(\prod_{i=1}^N g_i\right),
\quad
g_i
=
\frac{1}{\sqrt3}
\begin{pmatrix}
-|0\rangle_i & \sqrt2\,|+\rangle_i\\
-\sqrt2\,|-\rangle_i & |0\rangle_i
\end{pmatrix}.
\end{equation}

The transfer matrix is
\begin{equation}
E=\sum_{\sigma=+,0,-}C^\sigma\otimes C^\sigma
=
\begin{pmatrix}
1/3&0&0&2/3\\
0&-1/3&0&0\\
0&0&-1/3&0\\
2/3&0&0&1/3
\end{pmatrix}
=
\Pi_0+r\Pi_1,
\end{equation}
with eigenvalues $\lambda_0=1$ and $\lambda_1=\lambda_2=\lambda_3=r\equiv -1/3$. The normalized eigenvectors are
\begin{equation}
\begin{aligned}
|R_0\rangle&=|L_0\rangle
=\frac1{\sqrt2}(1,0,0,1)^T,\\
|R_1\rangle&=|L_1\rangle
=\frac1{\sqrt2}(1,0,0,-1)^T,\\
|R_2\rangle&=|L_2\rangle=(0,1,0,0)^T,\\
|R_3\rangle&=|L_3\rangle=(0,0,1,0)^T.
\end{aligned}
\end{equation}
In the compact notation we write $\Pi_0=|0\rangle\langle 0|$ with $|0\rangle=(1,0,0,1)^T/\sqrt2$, $\Pi_1=I_4-\Pi_0$, and introduce the auxiliary vector $|z\rangle=(1,0,0,-1)^T/\sqrt2$.

The elementary excitations (solitons or ``crackons'') are constructed by inserting a spherical Pauli matrix $\sigma^\mu$ ($\mu=0,\pm1$) at site $n$,
\begin{equation}
|C_n^\mu\rangle
=
\mathrm{Tr}\left[
\prod_{i=1}^{n-1}g_i\,(\sigma^\mu g_n)\prod_{i=n+1}^{N}g_i
\right],
\end{equation}
where $\sigma^0=\sigma_z=\mathrm{diag}(1,-1)$ and $\sigma^{\pm1}=\mp(\sigma_x\pm i\sigma_y)/\sqrt2$ are the Pauli matrices in the spherical basis. A momentum eigenstate is
\begin{equation}
|C^\mu(k)\rangle
=
\sum_{n=1}^N e^{ikn}|C_n^\mu\rangle.
\end{equation}
In the doubled (bra-ket) transfer space, the ket defect is generated by $P=\sigma_z\otimes I_2$ and the bra defect by $Q=I_2\otimes\sigma_z$; they satisfy $P|0\rangle=Q|0\rangle=|z\rangle$ and $P|z\rangle=Q|z\rangle=|0\rangle$.

\subsection{Excitation energy: per-site norm and numerator}
\label{sec:norm_num}

Because the Hamiltonian is rotationally invariant, the three branches $\mu=0,\pm1$ are exactly degenerate. We therefore suppress the index $\mu$ and define the excitation energy through the Rayleigh quotient
\begin{equation}
\epsilon^\mu(k)
=
\frac{\langle C^\mu(k)|(\hat H-E_0)|C^\mu(k)\rangle}
{\langle C^\mu(k)|C^\mu(k)\rangle}
\equiv
\frac{\mathcal M(k)}{\mathcal N(k)},
\end{equation}
where $E_0=N e_0$ is the total AKLT ground-state energy and $e_0$ is the corresponding energy density. Here we use the per-site norm and per-site numerator
\begin{equation}
\mathcal N(k)
=
\lim_{N\to\infty}\frac{1}{N}\langle C^\mu(k)|C^\mu(k)\rangle,
\qquad
\mathcal M(k)
=
\lim_{N\to\infty}\frac{1}{N}\langle C^\mu(k)|(\hat H-E_0)|C^\mu(k)\rangle .
\end{equation}
The overall factor $N$ cancels in the Rayleigh quotient. This per-site definition is convenient because, at finite $N$, the discrete sums differ from the infinite geometric series only by boundary terms of order $O(1/N)$, which vanish in the thermodynamic limit. We also introduce the two geometric series
\begin{equation}\label{SMEQ:GeoSeries}
\begin{aligned}
G(k)&=\sum_{d\in\mathbb Z}e^{ikd}r^{|d|}
=\frac{1-r^2}{1-2r\cos k+r^2}
=\frac{4}{5+3\cos k},\\
D(k)&=\sum_{d\in\mathbb Z}|d|\,e^{ikd}r^{|d|}=\frac{2r(1+r^2)\cos k-4r^2}{\left(1-2r\cos k+r^2\right)^2}
=-\frac{3(3+5\cos k)}{(5+3\cos k)^2},
\end{aligned}
\end{equation}
where $G$ denotes the geometric overlap sum and $D$ denotes the distance-weighted sum.

\subsubsection{Per-site norm $\mathcal N(k)$}

The overlap between two crackions separated by $d$ sites is
\begin{equation}
\langle C_0^\mu|C_d^\mu\rangle
=
\langle 0|P E^{d}Q|0\rangle
=
\langle z|E^{d}|z\rangle
=
r^{|d|},
\end{equation}
where the self-overlap $\langle 0|PQ|0\rangle=1$ has been used. Consequently,
\begin{equation}\label{SMEQ:NormNk}
\mathcal N(k)
=
\lim_{N\to\infty}\frac{1}{N}
\sum_{n,m=1}^N e^{ik(n-m)}r^{|n-m|}
=
\sum_{d\in\mathbb Z}e^{ikd}r^{|d|}
=
G(k),
\end{equation}
where $G(k)$ is given in Eq.~\eqref{SMEQ:GeoSeries}. The nontrivial denominator $5+3\cos k$ in $G(k)$ is a direct consequence of the non-orthogonality of the crackion position states, not of any particular interaction term.

\subsubsection{Per-site numerator $\mathcal M(k)$}

The AKLT reference energy density is obtained from the correlation functions
\begin{equation}
\langle\mathbf S_i\cdot\mathbf S_{i+1}\rangle=-\frac43,\qquad
\langle\mathbf S_i\cdot\mathbf S_{i+2}\rangle=\frac49,
\qquad
e_0=-\frac43+\frac49\alpha.
\end{equation}
We define the connected matrix elements
\begin{equation}
\tilde h_d
=
\lim_{N\to\infty}\langle C_0^\mu|(\hat H-N e_0)|C_d^\mu\rangle
=
h_d^{\mathrm{NN}}+\alpha\,h_d^{\mathrm{NNN}}.
\end{equation}
Because the subtraction of the reference energy $N e_0$ must be applied to the same non-orthogonal overlap $\langle C_0^\mu|C_d^\mu\rangle=r^{|d|}$, the subtraction is performed on the connected part rather than term by term. A direct tensor-network contraction then gives the short-range matrix elements
\begin{equation}
\tilde h_0=\frac{16}{9}-\frac{32\alpha}{27},\qquad
\tilde h_{\pm1}=-\frac{16\alpha}{27}.
\end{equation}
For $|d|\ge2$, the two defects are separated beyond the range of the Hamiltonian. The long-range matrix element contains both $r^{|d|}$ and $|d|\,r^{|d|}$:
\begin{equation}
\tilde h_d=\Bigl(A+\gamma|d|\Bigr)r^{|d|},\qquad |d|\ge2,
\end{equation}
with
\begin{equation}
A=\frac83-16\alpha,\qquad
\gamma=-\frac83+\frac{104\alpha}{9}.
\end{equation}
The factor $|d|$ arises from counting the bonds lying between the two defects: for $|d|\ge2$, the intermediate region contributes $|d|-1$ nearest-neighbor bonds (each with net coefficient $-8/3$) and $|d|-2$ next-nearest-neighbor bonds (each with $104\alpha/9$), together with two boundary terms. The threefold degeneracy of the transfer matrix alone does not generate the linear factor.

For the Fourier sum, it is convenient to write the matrix elements for all $d$ in the same form by adding local corrections for $d=0,\pm1$. We thus define
\begin{equation}
b_0=\tilde h_0-A=-\frac89+\frac{400\alpha}{27},\qquad
b_1=\tilde h_1-(A+\gamma)r=-\frac{56\alpha}{27},
\end{equation}
so that $\tilde h_d=(A+\gamma|d|)r^{|d|}$ holds for $|d|\ge2$, while $\tilde h_0=A+b_0$ and $\tilde h_{\pm1}=(A+\gamma)r+b_1$. Using the definitions of $G(k)$ and $D(k)$ in Eq.~\eqref{SMEQ:GeoSeries}, one obtains the per-site numerator
\begin{equation}\label{SMEQ:NormMk}
\mathcal M(k)
=
\sum_{d\in\mathbb Z} e^{ikd}\tilde h_d
=
A\,G(k)+\gamma\,D(k)+b_0+2b_1\cos k.
\end{equation}

\subsection{Final dispersion and physical consequences}
\label{sec:dispersion}

Dividing the per-site numerator $\mathcal M(k)$ by the per-site norm $\mathcal N(k)=G(k)$, the excitation energy $\epsilon^\mu(k)=\mathcal M(k)/\mathcal N(k)$ becomes
\begin{equation}
\begin{aligned}
\epsilon^\mu(k)
&=A+\gamma\,\frac{D(k)}{G(k)}
+\frac{b_0+2b_1\cos k}{G(k)}\\
&=A-\frac{3\gamma}{4}\frac{3+5\cos k}{5+3\cos k}
+\frac{5+3\cos k}{4}\left(b_0+2b_1\cos k\right).
\end{aligned}
\end{equation}
Substituting the explicit values of $A$, $\gamma$, $b_0$, and $b_1$, and using $2\cos^2 k=1+\cos 2k$, we arrive at the closed-form dispersion
\begin{equation}\label{SMEQ:DispRslt}
\boxed{
\begin{aligned}
\epsilon^\mu(k)
&=
\frac{14}{9}
+\frac{26}{27}\alpha
+\frac{160\alpha-18}{27}\cos k
-\frac{14}{9}\alpha\cos(2k)
+
\left(2-\frac{26\alpha}{3}\right)
\frac{3+5\cos k}{5+3\cos k},
\end{aligned}
}
\end{equation}
where $\mu=0,\pm1$ labels the three degenerate spin branches. This result is identical to Eq.~(18) of Ref.~\cite{SMKolezhuk1997PRB}.

For sufficiently large $\alpha$, the system decouples into two independent subchains. The excitation spectrum of a single subchain, obtained by keeping the other subchain in the AKLT state, is
\begin{equation}
\epsilon_{\mathrm{NNN-AKLT}}(k)
=
\alpha\,\epsilon_{\mathrm{AKLT}}(2k;\alpha=0),
\end{equation}
with a gap at $k=\pi/2$ equal to $2\alpha/9$. Equating the ground-state energies of the two competing variational states, $E_0^{\mathrm{AKLT}}(\alpha)=E_0^{\mathrm{NNN-AKLT}}(\alpha)$, locates the variational transition point at
\begin{equation}
\alpha_T^{\mathrm{var}}=\frac34=0.75.
\end{equation}
At this value, we determine the minimum of the variational dispersion Eq.~\eqref{SMEQ:DispRslt} by minimizing it for $0\le k\le\pi$. Setting $c=\cos k$, the stationary-point condition reduces to the cubic equation $189c^3+477c^2+15c-101=0$. The physical solution is $c\approx -0.5393$, corresponding to $k_\ast\approx 2.1404\approx 0.6813\pi$. Substituting this value back into Eq.~\eqref{SMEQ:DispRslt} yields the variational gap $\Delta^{\mathrm{var}}\approx 0.3246$, in good agreement with the value of approximately $0.325$ reported in the original paper, and consistent with a first-order transition with nonzero excitation gap.
Thus the variational MPS approach, despite its simplicity, captures the essential physics of the frustrated $S=1$ chain and provides a solid analytical foundation for the numerical observations reported in the main text.

\begin{figure}[!ht]
\centering
\includegraphics[width=0.50\columnwidth, clip]{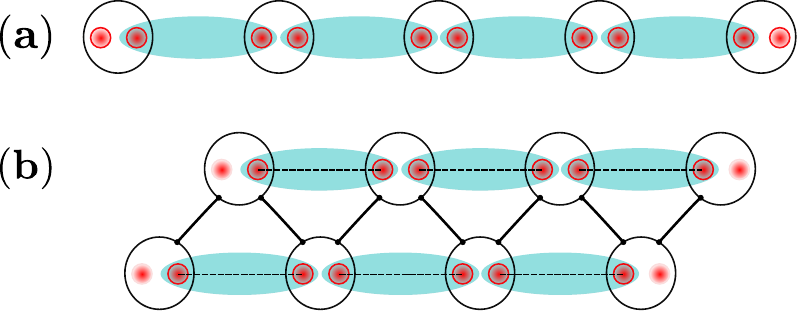}\\
    \caption{Schematic representation of (a) Haldane phase (a.k.a. AKLT phase) and (b) NNN-Haldane phase (a.k.a. NNN-AKLT phase), characterized by dominant nearest-neighbor and next-nearest-neighbor singlet bonds, respectively. Solid lines trace the chain paths, and shaded regions indicate singlet bonds.
    }\label{FIG:HDVBS}
\end{figure}

\section{Difference of Bond Strength} \label{SEC:AppA}

The valence-bond representations of the Haldane and NNN-Haldane phases are illustrated in Figs.~\ref{FIG:HDVBS}(a) and (b), respectively. Depending on the degree of frustration, the chain develops a dominant nearest-neighbor or next-nearest-neighbor valence-bond pattern. It is therefore natural to introduce the difference of bond strength (DBS),
\begin{equation}
\mathcal{D}_{L} = \big|\langle S_{L/2}^z S_{L/2+2}^z\rangle - \langle S_{L/2}^z S_{L/2+1}^z\rangle\big|,
\end{equation}
as a diagnostic tool for identifying the first-order transition point \cite{SMLuo2019PRB}.

\begin{figure}[!ht]
\centering
\includegraphics[width=0.45\columnwidth, clip]{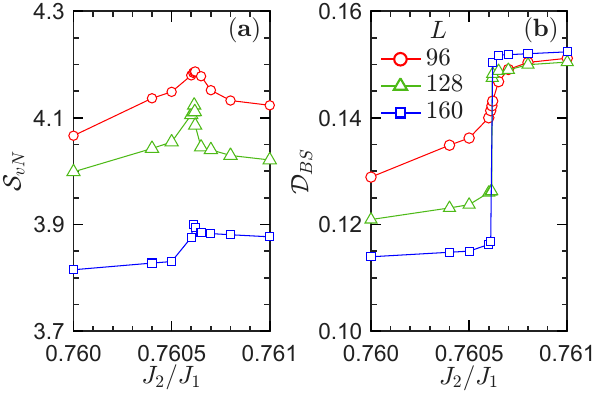}\\
    \caption{(a) von Neumann entropy $\mathcal{S}_{vN}$ and (b) difference of bond strength $\mathcal{D}_{BS}$ as functions of $J_2/J_1$ for the $J_1$-$J_2$ spin-1 chain with $L = 96$ (red circles), $128$ (green triangles), and $160$ (blue squares) under periodic boundary conditions.
    }\label{FIG:1stDBS}
\end{figure}

Consistent with the high-precision transition point obtained from the iDMRG calculations of the von Neumann entropy and the DMRG calculations of the two string order parameters in the main text, the abrupt changes in the von Neumann entropy and the DBS shown in Figs.~\ref{FIG:1stDBS}(a) and (b) unambiguously pinpoint $J_{2,t} = 0.760616(2)$. These results demonstrate that the bond-reversal method not only captures the hallmark discontinuity of the weakly first-order transition but also serves as a versatile probe that does not rely on detailed knowledge of the underlying physical properties \cite{SMLuo2019PRB}.

\begin{figure}[!ht]
\centering
\includegraphics[width=0.45\columnwidth, clip]{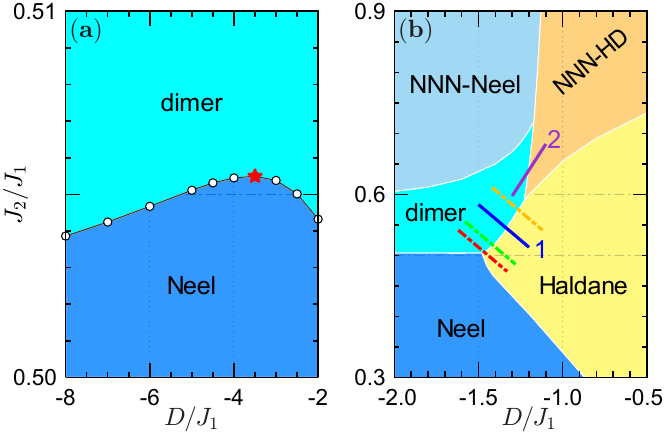}\\
\caption{(a) Phase boundary of the dimer--N{\'e}el transition at large $|D|$ case.
    The red pentagram marks the peak of the locus.
    (b) Two cutting lines of $J_2$ = 7/30 ($\alpha-D$) (1, red) and $J_2 = 0.425D + 1.15$ (2, blue).
    }\label{FIG:SubPDNegD}
\end{figure}

\begin{figure}[!ht]
  \centering
  \includegraphics[width=0.45\columnwidth, clip]{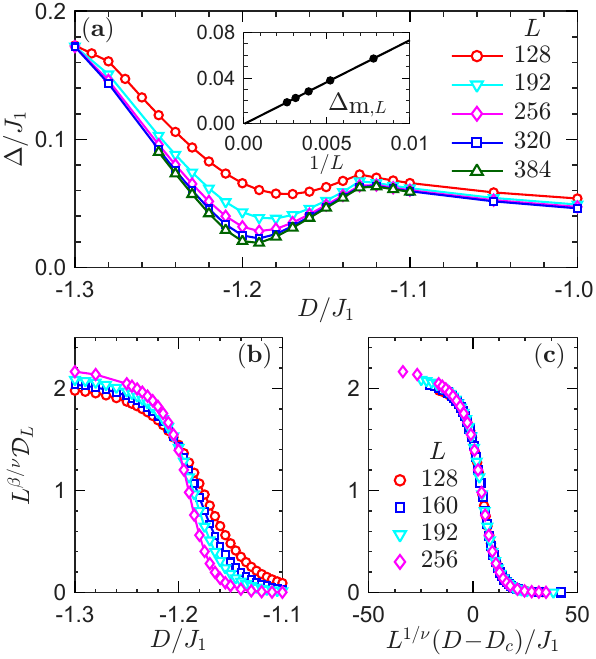}\\
  \caption{(a) Lowest excitation gap $\Delta_1$ as a function of $D/J_1$ along the cut $J_2 = 0.425D + 1.15$ for chain lengths $L = 128$ to $384$. The inset shows the finite-size extrapolation of the minimum gap near the transition.
  (b), (c) Finite-size scaling analysis of the dimer order parameter $\mathcal{D}_L$ along the same cut, yielding the critical point $D_c = -1.2020(5)$ and critical exponents $\beta = 0.130(7)$ and $\nu = 0.96(5)$.}
  \label{FIG:DimerNNNHDGAP}
\end{figure}

\section{Non-monotonic Boundary and Characteristic Cuts} \label{SEC:AppB}

Figure~\ref{FIG:SubPDNegD} presents the quantum phase diagram centered on the dimerized phase. Fig.~\ref{FIG:SubPDNegD}(a) reveals that the phase boundary between the dimerized and N\'eel phases exhibits a non-monotonic shape as the negative SIA intensity increases, featuring a distinct hump near $D/J_1 \approx -3.5$. Fig.~\ref{FIG:SubPDNegD}(b) displays several representative cuts starting from within the dimerized phase. The transition to the NNN-Haldane phase is probed along \textbf{Cut II}, defined by $J_2 = 0.425D + 1.15$. \textbf{Cut I} explores the transition to the Haldane phase and consists of four segments, parameterized as $J_2 = \frac{7}{30}(\alpha - D)$ with $\alpha = 0.7$ (red), $0.8$ (green), $1.0$ (blue), and $1.2$ (yellow).

We first examine the transition between the dimerized and NNN-Haldane phases along Cut II, defined by $J_2 = 0.425D + 1.15$, which traverses the phase boundary. Given that both phases possess unique ground states under open boundary conditions, we employ the first excitation gap $\Delta_1 = E_1 - E_0$ to locate the transition. As shown in Fig.~\ref{FIG:DimerNNNHDGAP}(a), $\Delta_1$ exhibits nonmonotonic behavior as $D$ varies. Both phases are gapped, with $\Delta_1$ remaining stable away from the critical region; near the transition, however, it decreases systematically with increasing chain length $L$. A linear extrapolation of the minimum of $\Delta_1$ near the dip (inset) yields $0.0003(7)$, consistent with gap closure across the quantum phase transition.

Finite-size scaling analysis of the order parameter along Cut II yields the critical point $D_c = -1.2020(5)$ and critical exponents $\beta = 0.130(7)$ and $\nu = 0.96(5)$ [see Figs.~\ref{FIG:DimerNNNHDGAP}(b) and (c)], placing the transition squarely in the Ising universality class.

\end{CJK*}

\end{document}